\documentclass[preprint,showpacs,preprintnumbers,amsmath,amssymb]{revtex4}

\usepackage{graphicx}
\usepackage{dcolumn}
\usepackage{bm}
\usepackage{amsmath,dsfont}
\usepackage{amssymb,amsthm,amscd, amsbsy, array}
\usepackage{graphics,graphicx,xcolor}
\usepackage{etex}
\usepackage[all]{xy}

\numberwithin{equation}{section}

\newcommand{\half}{{\scriptstyle{\frac{1}{2}}}}
\def\2{{\half}}
\def\smallcirc{{\,\raise 0.5pt \hbox{$\scriptstyle\circ$}\,}}
\newcommand{\const}{\mathop{\rm const}\nolimits}
\def\parag{\hfil\break} 
\def\kikezd{\parag\underbar}

\def\p{{\partial}}

\def\bE{{\bf E}}

\newcommand{\vB}{{\bm{B}}}
\newcommand{\bA}{{\bm A}}

\newcommand{\vC}{{\bm C}}

\newcommand{\vp}{{\bm p}}
\newcommand{\bp}{{\bm p}}
\newcommand{\Orth}{{\rm O}}

\def\br{{\bm{r}}}
\def\bK{{\bm{K}}}
\def\bL{{\bm{L}}}
\def\beq{\begin{equation}}
\def\eeq{\end{equation}}
\def\beqa{\begin{eqnarray}}
\def\eeqa{\end{eqnarray}}

\def\barray{\left(\begin{array}}
\def\earray{\end{array}\right)}
\def\barraynb{\begin{array}}
\def\earraynb{\end{array}}

\def\smallover#1/#2{\hbox{$\textstyle\frac{#1}{#2}$}} %

\def\vx{{\bm{x}}}

\def\vnabla{{\overrightarrow{\nabla}}}

\newcommand{\vK}{{\bm K}}

\newtheorem{theorem}{Theorem}
\def\bec{\begin{center}}
\def\ec{\end{center}}
\def\bequ{\begin{enumerate}}
\def\eequ{\end{enumerate}}

\usepackage{color}
\newcommand{\red}{\textcolor{red}}  
\newcommand{\blue}{\textcolor{blue}}
\newcommand{\purple}{\textcolor{purple}}

\newcommand{\cyan}{\textcolor{cyan}}
\newcommand{\magenta}{\textcolor{magenta}}

\begin{document}

\preprint{arXiv:1308.3035v3
}

\title{Separability and
Dynamical Symmetry of Quantum Dots
\\[6pt]
}

\author{
P.-M. Zhang$^{1}$\footnote{e-mail:zhpm@impcas.ac.cn},
L.-P. Zou$^{1}$\footnote{e-mail:zoulp@impcas.ac.cn},
P.~A.~Horvathy$^{1,2}$\footnote{
e-mail:horvathy-at-lmpt.univ-tours.fr},
G.~W.Gibbons$^{3}$\footnote{mail:G.W.Gibbons@damtp.cam.ac.uk},
}

\affiliation{$^{1}$Institute of Modern Physics, Chinese Academy of Sciences
\\
Lanzhou (China)
\\
$^{2}${\it Laboratoire de Math\'ematiques et de Physique Th\'eorique}, Tours University
(France).
\\
$^{3}$Department of Applied Mathematics and Theoretical  Physics,
Cambridge University, Cambridge, UK\\
}

\date{\today}

\begin{abstract}The separability and Runge-Lenz-type dynamical symmetry of the internal dynamics of certain two-electron Quantum Dots, found by Simonovi\'c et al. [1], is traced back to that of the perturbed Kepler problem. A large class of axially symmetric perturbing potentials which allow for separation in parabolic coordinates can easily be found. Apart of the 2:1 anisotropic harmonic trapping potential considered in [1], they include a constant electric field parallel to the magnetic field (Stark effect), the ring-shaped Hartmann potential, etc. The harmonic case is studied in detail.
\end{abstract}


\noindent
 Annals of Physics {\bf 347}, 94 -- 116 (2013)\\
 http://dx.doi.org/10.1016/j.aop.2013.11.004

\maketitle

\tableofcontents


\section{Introduction}

A two-electron quantum dot (QD) in a perpendicular magnetic field, described by the Hamiltonian,
\beq
H=\sum_{a=1}^2\left[\frac{1}{2M}\left(\vp_a-e\bA_a\right)^2+U(\br_a)\right]-\frac{a}{|\br_1-\br_2|}\,,
\label{fullHam}
\eeq
where the confining potential is that of an axially symmetric oscillator \cite{Simon,Simon2},
\beq
U(\br)=\frac{M}{2}\left[\omega_0^2\big(x^2+y^2\big)
+\omega_z^2z^2\right],
\eeq
may carry unexpected symmetries.
Firstly, the system splits, consistently with 
Kohn's theorem, into center-of-mass and relative motion and the former system carries a Newton-Hooke type symmetry \cite{Kohn,ZHAGK}.
Secondly, for the particular values of the frequency ratios
\beq
\tau=\frac{\omega_z}{\sqrt{\omega_0^2+\omega_L^2}}=1,\,2,
\label{parttau}
\eeq
where $\omega_L$ is the Larmor frequency \footnote{In the QD problem the Larmor frequency  involves  the reduced mass
$M^*=M/2$, $\omega_L=eB/2M^*$.},
the \emph{relative} motion becomes separable in suitable coordinates \cite{Simon}, which hints at further symmetry.
This paper is devoted to the study of the latter,
and to generalizing them to other axi-symmetric trapping potentials.
 
Our first step is to trace back the problem to those results found earlier for a particle without a magnetic field, $\vB=0$ \cite{Alhassid,Blumel}.
%
%
 Choosing the vector potential $\bA=\half B(-y,x,0)$ and 
introducing  $\bm{R}=(\br_1+\br_2)/2$ and
$\br=\br_1-\br_2$, the system splits into 
center-of-mass and relative parts. 
Disregarding the first, we focus our attention at the relative motion.
Following \cite{Simon}, the relative  Hamiltonian becomes, after suitable re-definition, 
\beq
H\equiv H_{rel}=-\frac{1}{2M^*}\big(\vnabla_\rho-eiA_\rho\big)^2+\frac{M^*}{2}\Big(
\omega_0^2(x^2+y^2)+\omega_z^2z^2\Big)-\frac{a}{r},
\label{relHam}
\eeq
where $M^*=M/2$ is the reduced mass
and we used units where $\hbar=1$.
Now   putting 
\beq
\br\to R(t)\,\br, \qquad
R(t)=\barray{cc}
\cos\omega_L\, t &\sin\omega_L\, t
\\[4pt]
-\sin\omega_L\, t &\cos\omega_L\, t
\earray,
\qquad
\omega_L=\frac{eB}{2M^*}
\label{B-O}
\eeq
eliminates the vector potential altogether and 
the Schr\"odinger equation of relative motion,
$\big[i\p_t-H_{rel}\big]\psi=0$, goes over into 
\begin{equation}
\left[i\partial_t+\underbrace{\,\frac {\bigtriangleup}{2}+\frac{a}{r}\,}_{Kepler}-
\underbrace{\frac
12(\omega_0^2+\omega_L^2)\left(x^2+y^2\right)-\frac 12\omega
_z^2z^2}_{axi-symmetric\; oscillator}\right]\psi=0,
\label{pertKpb}
\end{equation}
where we also assumed that $M^*=1$.

The rotational trick (\ref{B-O}) allowed us, hence, to convert the constant-magnetic-field problem into that of \emph{the Kepler potential perturbed by an axially symmetric oscillator} \cite{Alhassid,Blumel}.
 In what follows, we only study the latter problem, since all results can be translated to the constant-magnetic context by applying (\ref{B-O}) backwards. Note that in the original QD problem the electrons repel and thus $a\propto -e^2<0$; 
 for completeness, we also consider here the attractive Kepler case $a>0$. 
Our analysis bears also strong similarities with that of ions in a
Paul trap \cite{Blumel}.

\section{Classical separability}\label{3Dseparability}

We first study the
 classical context, where ``separability'' refers to that of the Hamilton-Jacobi equation.
According to the Robertson Theorem (\cite{Cordani} (Sec. 8.1.3., p. 169), see also
 \cite{Benenti}), classical separability does imply, in our case, that of the Schr\"odinger equation \ref{Quantum}.
Restricting ourselves to 
natural orthogonal systems, i.e., such whose Hamiltonian is 
\begin{equation}
H=\frac{1}{2}\sum_{k=1}^{n}g_{k}(x_{1},\ldots
,x_{n})\,p_{k}^{2}+V(x_{1},\ldots ,x_{n}),
\label{natsyst}
\end{equation}%
the answer is given by~:
\begin{theorem}[\emph{St\"{a}ckel}  \cite{Cordani}]

An\nolinebreak\ $n$-dimensional system with Hamiltonian (\ref{natsyst})
 is separable if and only if there
exists 
(i)
an invertible $n\times n$ matrix and
(ii) a column vector, 
\beq
(i)\;\;
\text{\textbf{\textsf{U}}}=
\barray{ccccc}
U_{11}&\dots &U_{1k} &\dots &U_{1n}
\\
\vdots &\vdots &\vdots &\vdots &\vdots
\\ 
U_{n1}&\dots &U_{nk} &\dots &U_{nn}
\\
\earray
\qquad\hbox{and}\qquad
(ii)\;\;\underline{w}=\left( 
\begin{array}{c}
w_{1} \\ 
\vdots \\ 
w_{n}%
\end{array}%
\right),
\label{Stackelmatrixvector}
\eeq
called the \emph{St\"{a}ckel matrix} and the 
\emph{St\"{a}ckel vector}, respectively, 
 whose $j$-th rows are functions of $x_{j}$
only, and such that%
\begin{equation}
\sum_{j=1}^{n}g_{j}U_{jk}=\delta_{1k},\qquad \sum_{j=1}^{n}g_{j}w_{j}=V.
\label{StackelHypotheses}
\end{equation}
\end{theorem}

\vskip2mm
That the St\"ackel conditions are  necessary is proved in Ref. \cite{Cordani}.
Here we only show how to use them.
Put%
\begin{equation} 
\underline{p}^{2}=\left( 
\begin{array}{c}
p_{1}^{2} \\ 
\vdots \\ 
p_{n}^{2}%
\end{array}%
\right) ,
\qquad 
\underline{\alpha }=\left( 
\begin{array}{c}
\alpha_{1} \\ 
\vdots \\ 
\alpha_{n}%
\end{array}%
\right),
\end{equation}%
where the $\alpha_i$s are arbitrary constants,
and define the column vector $\underline{K}$
composed of $n$ functions,%
\begin{equation}
\underline{K}(x_{j},p_{k})=\text{\textbf{\textsf{U}}}^{-1}\Big(\frac{1}{2}\underline{p}%
^{2}+\underline{w}\Big).  
\label{first_int}
\end{equation}%
Note for further record that, owing to (\ref{StackelHypotheses}), the first of these functions is in fact the Hamiltonian. 
Then the Hamil\-ton--Jacobi Equation can be viewed as the first row of
the system of $n$ equations 
\beq
\underline{K}(x_{j},p_{k})=\, \underline{\alpha}.
\eeq 
Inverting this  relation,
$
\frac{1}{2}\underline{p}^{2}+\underline{w}=\text{\textbf{\textsf{U}}}%
\underline{\alpha },
$
defines $p_{k}$ implicitly as a function of the $x_{k}$ and of the constants $\alpha_{1},\ldots
, \alpha_{n}.$ Putting $\frac{\partial S_{k}}{\partial x_{k}}=p_{k},$ we see
that 
$ S=\sum_{k}S_{k},
\;
S_{k}=S_{k}(x_{k},\alpha_{1},\ldots,\alpha_{n})
$ 
is a complete integral. $S$ is in fact a solution of the Hamilton--Jacobi Equation by
construction, and
one readily shows that $\det\left( \frac{\partial^{2}S}{\partial \alpha_{j}\partial x_{k}}\right)\neq0$, cf. \cite{Cordani}.

The $n$ functions $\underline{K}(x_{j},p_{k})$ are  first integrals
in involution; they are quadratic in the momenta and, in coordinates
allowing for separation, they do not contain products of the momenta. Our
problem is precisely to find such coordinate systems, and the Eisenhart Theorem \cite{Eisenhart}
 (\cite{Cordani}
chapter 8) provides us with a constructive method for doing this. 

Turning to our concrete problem here,
let us first remind the reader that
the unperturbed Kepler Hamiltonian,%
\begin{equation}
H_{\mathrm{Kepler}}=\frac{1}{2}\bp^{2}-\frac{a}{r},
\label{KeplerHam}
\end{equation}%
is separable in four coordinate systems, namely in spherical, (semi)parabolic,
elliptic and spheroconical ones \cite{Cordani}.  

Turning to the QD problem which is our main interest here, the relative Hamiltonian $H\equiv H_{rel}$ reads, after elimination of the magnetic field by switching to rotating coordinates,  the Kepler 
 problem perturbed by a harmonic (but not necessarily isotropic) oscillator, 
\begin{equation}
H=H_{\mathrm{Kepler}}+V,
\qquad
V=V_{osc}=\frac{1}{2}\Big(\omega_{\rho}^{2}\rho^{2}+\omega_{z}^{2}z^{2}\Big),
\label{perturbedKepler}
\end{equation}%
where $\rho=\sqrt{x^2+y^2}$,  $\omega_\rho=\sqrt{\omega_L^2+\omega_0^2}$
cf. (\ref{pertKpb}),
and inquire about the values of the parameters $\omega_{\rho}$ and $\omega_{z}$ that make $H$ separable in one or another of  the four ``good'' coordinate systems mentioned above.

\begin{figure}
\begin{center}
\includegraphics[scale=.65]{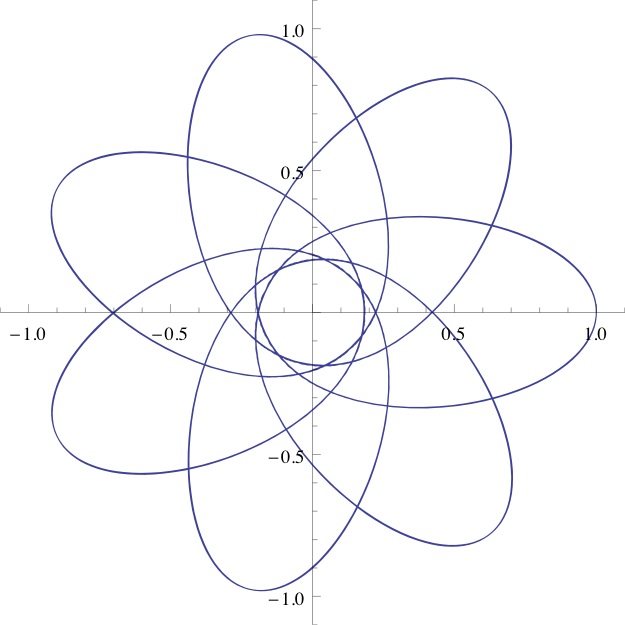}
\vspace{-8mm}
\end{center}
\caption{\it The Kepler problem combined with an isotropic oscillator is separable. The trajectories are perturbed Kepler ellipses rotating around in the plane perpendicular to the angular momentum.
}
\label{KeplerOsc11}
\end{figure}

$\bullet$ In the spherical case things are simple and do not require any
calculation, and we only mention it for pedagogical purposes. For $\omega_{\rho}=\omega_{z}=\omega$ the perturbation we added is itself
isotropic and the Hamiltonian is plainly separable in spherical coordinates. For completeness and for further use, we record the
St\"ackel matrix and  ector, respectively,
\begin{eqnarray}
\text{\textbf{\textsf{U}}} &=&\left( 
\begin{array}{ccc}
1& -\frac{1}{r^2}&0 
\\ 
0&1&-\frac{1}{\sin^2\theta} %
\\
0&0&1
\end{array}%
\right)
\quad\Rightarrow\quad
\text{\textbf{\textsf{U}}}^{-1}=\left( 
\begin{array}{ccc}
1 & \frac{1}{r^2}&\;\frac{1}{r^2\sin^2\theta}
\\ 
0&1&\frac{1}{\sin^2\theta}
\\
0&0&1
\end{array}%
\right) , 
\\[12pt]
\underline{w} &=&\left( 
\begin{array}{c}
-\frac{a}{r}\\ 
0\\
0%
\end{array}%
\right)
+\left(\begin{array}{c}
\frac{\omega^2}{2}r^2 \\ 
0\\
0%
\end{array}%
\right).
\label{3DsphStaeckel}
\end{eqnarray}%
The three commuting conserved quantities in involution mentioned above are, therefore, (i) the Hamiltonian, (ii) the half of the square of the total angular momentum, $L^2/2$, and (iii) the half of the squared $z$-component of the angular momentum, $L_z^2/2$, associated with the 
rotational $\Orth(3)$ symmetry --- generalizing the pure Kepler problem \cite{Cordani}.
Here we do not pursue this issue and merely plot some  trajectories, see Fig.
\ref{KeplerOsc11}.

$\bullet$ The (semi)parabolic case, which is our main concern in this paper, with
 coordinates $(\xi\geq0,\eta\geq0,2\pi\geq\varphi\geq0)$,
\beq
x=\xi\eta\,\cos\varphi,
\quad
y=\xi\eta\,\sin\varphi,
\quad
z=\frac12(\xi^2-\eta^2),
\label{sparabc}
\eeq
is non-trivial, though. 
The St\"{a}ckel matrix and vector read, respectively,
\begin{eqnarray}
\text{\textbf{\textsf{U}}} &=&\left( 
\begin{array}{ccc}
\xi^{2} & -1&- \frac{1}{\xi^2} 
\\ 
\eta^{2} &1&- \frac{1}{\eta^2} %
\\
0&0&1
\end{array}%
\right) 
\quad\Rightarrow\quad
\text{\textbf{\textsf{U}}}^{-1}=\frac{1}{\xi^{2}+\eta^{2}}\left( 
\begin{array}{ccc}
1 & 1&\;\frac{1}{\xi^2}+\frac{1}{\eta^2}
\\ 
-\eta ^{2} & \xi^{2}&\;\frac{\xi^2}{\eta^2}-\frac{\eta^2}{\xi^2}
\\
0&0&\;\xi^2+\eta^2
\end{array}%
\right) , 
\\[12pt]
\underline{w} &=&\left( 
\begin{array}{c}
-a \\ 
-a\\
0%
\end{array}%
\right) +\left( 
\begin{array}{r}
f(\xi) \\ 
g(\eta)\\
h(\varphi)%
\end{array}%
\right) ,
\label{3DparaStaeckel}
\end{eqnarray}%
where $f(\xi)$ and $g(\eta)$ and $h(\varphi)$ are  arbitrary functions. Assuming axial symmetry, $h(\varphi)=0$.

Then our clue is that for the perturbed Kepler problem (\ref{perturbedKepler}) the St\"ackel condition is satisfied 
when the first row in
Eqns (\ref{first_int}) holds, and this happens 
\emph{whenever 
the perturbing potential}  satisfies 
\begin{equation}
(\xi^{2}+\eta^{2})V(\xi,\eta)=f(\xi)+g(\eta).
\label{seppar}
\end{equation}%
This simple but powerful \emph{separability condition} will lead to large classes of separable potentials, see Sec. \ref{seppot}. 
For our anisotropic oscillator,  it requires,
\begin{equation*}
(\xi^{2}+\eta^{2})V_{osc}(\xi,\eta)=\frac{1}{2}\left(\frac{\omega_{z}}{2}\right)^{2}(\xi
^{6}+\eta^{6})+\frac{1}{2}\left[\omega_{\rho}^{2}-\left(\frac{\omega
_{z}}{2}\right)^{2}\right]\Big(\xi^{4}\eta^{2}+\xi^{2}\eta^{4}\Big).
\end{equation*}%
Separability is hence achieved when\begin{equation}
\omega_{z}=2\omega_{\rho}
\quad\hbox{i.e.,\; for}\quad \tau=2.
\label{2:1}
\end{equation}%

Those three commuting conserved quantities in (\ref{first_int}) then read
\beqa
H&=&\underbrace{\displaystyle\frac1{2(\xi^2+\eta^2)}\left[
p_\xi^2+p_\eta^2
+\left(\frac1{\xi^2}+\frac1{\eta^2}\right)p_\varphi^2\right]-\frac{2a}{\xi^2+\eta^2}}_{Kepler\, Hamiltonian}\;+\;
\underbrace{\frac{\omega_\rho^2}{2}\left(\xi^4-\xi^2\eta^2+\eta^4\right)
}_{V_{osc}\;\;{with}\;\;\tau=2},\qquad
\label{KeposciHam}
\\[12pt]
K_z&=&\displaystyle\frac1{2(\xi^2+\eta^2)}\left[
\xi^2p_\eta^2-\eta^2p_\xi^2
+\left(\frac{\xi^2}{\eta^2}-\frac{\eta^2}{\xi^2}\right)p_\varphi^2\right]-a\,\frac{\xi^2-\eta^2}{\xi^2+\eta^2}
-
\frac{\omega_\rho^2}{2}\xi^2\eta^2(\xi^2-\eta^2),\qquad
\label{QDRLpar}
\\[8pt]
L_z^2/2&=&\frac12p_{\varphi}^2,
\eeqa
where
$p_\xi=(\xi^2+\eta^2)\,\dot{\xi},
\,
p_\eta=(\xi^2+\eta^2)\,\dot{\eta}$.
Translating into more familiar form,
\beqa
H&=&\displaystyle\frac{\vp^2}{2}
-\frac{a}{r}+V_{osc},
\\[4pt]
K_z&=&\underbrace{(\bp\times\bL)_z\,-\,a\frac{z}{r}}_{Kepler\, Runge-Lenz}
\;-\;
\omega_\rho^2\,\rho^2z,
\label{QDRLord}
\\
L_z^2/2&=&\frac12(\rho\dot{\varphi})^2,
\eeqa
allows us to interpret these quantities~:
 (i) $H$ is the perturbed Hamiltonian (\ref{perturbedKepler}), as it should;
(ii) $K_z$ 
 generalizes the \emph{$z$ component of the Runge-Lenz vector} and 
is indeed the separation constant found  in  \cite{Simon}.
The additional term $-\omega_\rho^2\rho^2z$  arises  due to the perturbing oscillator potential. 
 (iii)
The third quantity is, once again, the half of the \emph{squared
$z$ component of the angular momentum}. 
 The familiar Keplerian quantities \cite{Cordani} and those of the $2:1$ anisotropic oscillator \cite{Makar,Boyer} are recovered when $V_{osc}=0$ or when the Kepler potential is switched off, $a=0$, respectively.
Some classical trajectories will be presented in Sect. \ref{planarKO}.

\section{Reduction to and induction from the 2D problem}\label{planarKO}

Returning to classical aspects, let us observe
that the condition 
\beq
L_z\equiv p_\varphi=0
\label{Lzconstraint}
\eeq
constrains the motion into a ``vertical'' plane through the $z$ axis and in fact reduces the problem to the perturbed Kepler problem in 2D. 
Our strategy, in this Section,  will be to work backwards, starting with the 2D case and then extending to 3D.
 Putting $\varphi=0$ (say)
into the formulas in Section 
\ref{3Dseparability} provides us with two-dimensional ones. (\ref{sparabc}) yields, in particular, (semi)parabolic coordinates in the $x-z$ plane,
\beq 
x=x_+=\xi\eta,\qquad
z=\frac{1}{2}(\xi^{2}-\eta^{2}).
\label{2Dparabcoord}
\eeq
A subtlety arises, though: (\ref{2Dparabcoord}) is in fact  only \emph{half} of a coordinate system, since necessarily $x_+>0$, and should therefore be supplemented with  $x_-= - \xi\eta$ to cover the whole vertical plane. This problem is not present in 3D, since the first coordinate is indeed $\rho>0$, and the angular variable $\varphi$ takes care of the $x<0$ half plane, namely for
 $\varphi=\pi$.

The 2D St\"{a}ckel matrices and resp. vector are simply those in (\ref{3DparaStaeckel}) with the irrelevant $\varphi$-columns and rows erased.
 For our 2D anisotropic oscillator, separability is hence achieved for
\beq
\tau=\frac{\omega_{z}}{\omega_{\rho}}=2,
\eeq
just like before in 3D, cf. (\ref{2:1}).
Our theory provides us now with $D=2$ 
conserved quantities in involution, namely 
with the separable 2D Hamiltonian,  
\beq
H^0\equiv{H}\big|_{\varphi=0}=\underbrace{
\frac1{2(\xi^2+\eta^2)}\big(p_\xi^2+p_\eta^2
\;-\;4a\big)
}_{2D\ Kepler\ Hamiltonian }
+\frac{\omega_\rho^2}{2(\xi^2+\eta^2)}\big(\xi^6+\eta^6\big),
\label{xietaKOham}
\eeq
and with the Runge-Lenz-type conserved quantity 
\beq
K_z^0\equiv
K_z\big|_{\varphi=0}=\underbrace{\displaystyle\frac1{2(\xi^2+\eta^2)}\left[
\xi^2p_\eta^2-\eta^2p_\xi^2
 \right]-a\,\frac{\xi^2-\eta^2}{\xi^2+\eta^2}}_{2D\ Kepler\ Runge-Lenz-type}
\;-\;
{\omega_\rho^2}\underbrace{(\xi^2\eta^2)(\frac{\xi^2-\eta^2}{2})}_{\rho^2z}\,.
\label{QDRLphi0}
\eeq
 cf. (\ref{QDRLpar}).

\kikezd{More symmetries}

The unperturbed 2D Kepler problem has long been known to have an $\Orth(3)$ dynamical symmetry,  generated by the two components of the
Runge-Lenz vector, $\vK=(K_x,K_z)$, and by the
angular momentum, $L\equiv L_y$ perpendicular to the  $x-z$ plane \cite{JauchHill,CMI}.  In (semi)parabolic coordinates (\ref{2Dparabcoord}),
\beqa
K_x&=&\frac{1}{\sqrt{-2E}}\Big(p_\xi p_\eta-2E\,\xi\eta\Big),
\label{2DKx}
\\[6pt]
K_z&=&\frac{1}{\sqrt{-2E}}\Big(
\frac{p_\xi^2-p_\eta^2}{2}-2E\,\frac{\xi^2-\eta^2}{2}\Big),
\label{2DKz}
\\[6pt]
L&=&\frac12\Big(\eta p_\xi-\xi p_\eta\Big),
\label{2DAM}
\eeqa
where $E$ is a fixed value of the Kepler energy 
\beq
H_{Kepler}=\frac1{2(\xi^2+\eta^2)}\left(p_\xi^2+p_\eta^2
\;-\;4a\right),
\label{2DKH}
\eeq 
which is in fact the first term in (\ref{xietaKOham}),
as anticipated. 
Putting $H_{Kepler}$
 into (\ref{2DKz}) yields (\ref{QDRLphi0}) with $\omega_z=0$. The expression  (\ref{QDRLphi0}) generalizes, hence, the $z$-component of
the Runge-Lenz vector in the vertical plane, as anticipated.

Adding now, still in 2D, a perturbing oscillator potential to our pure Kepler problem destroys
\emph{most} of these symmetries. \emph{Most, but not all}, though~: the planar rotational symmetry generated by $L$ is  plainly broken by the anisotropy, but, for $\tau=2$,  the  corrected version (\ref{QDRLphi0}) of $K_z$ survives the perturbation. Numerical evidence also confirms that $K_x$ is also broken, except for $\tau=1/2$.

\begin{figure}
\includegraphics[scale=.34]{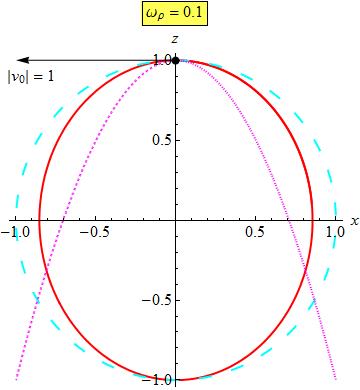}
\includegraphics[scale=.34]{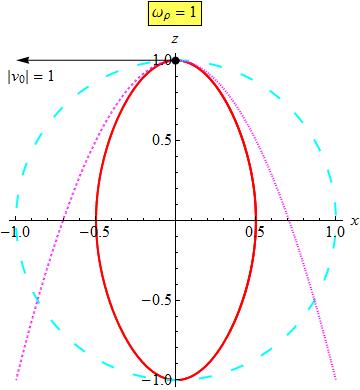}
\includegraphics[scale=.34]{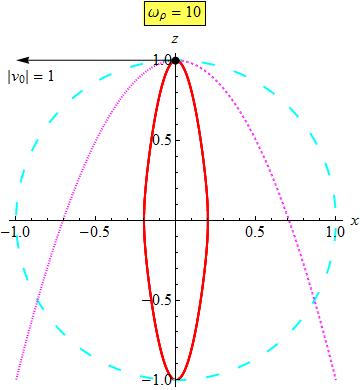}
\\[16pt]
\includegraphics[scale=.34]{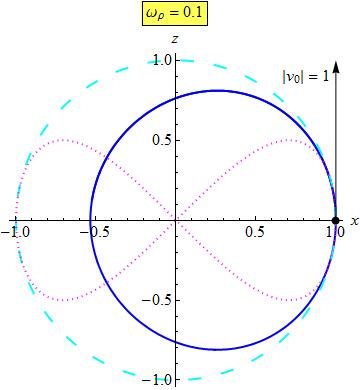}
\includegraphics[scale=.34]{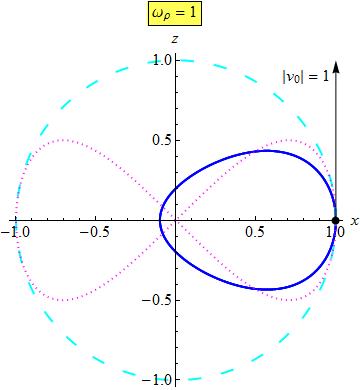}
\includegraphics[scale=.34]{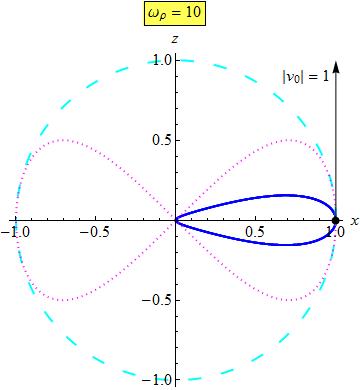}
\\[6pt]
\null\hskip-6mm
\includegraphics[scale=.44]{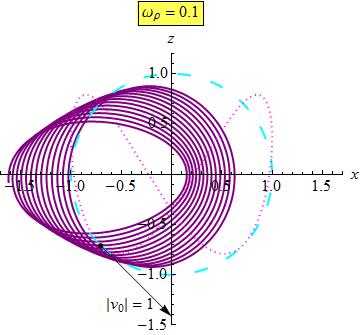}
\includegraphics[scale=.44]{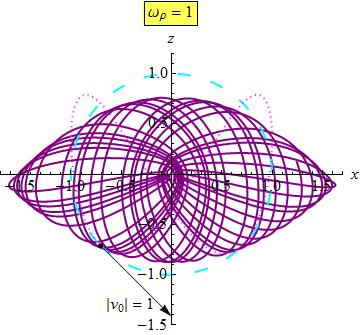}
\includegraphics[scale=.44]{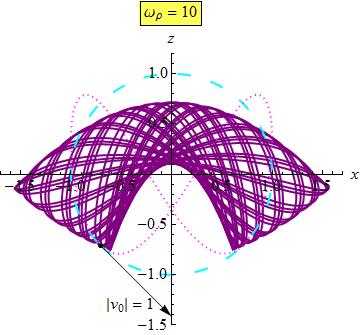}
\vspace{-6mm}
\caption{\it Trajectories in the classical planar Kepler problem perturbed with an axi-symmetric oscillator with anisotropy
$\tau=\omega_z/\omega_\rho=2$.
All figures have the same circular 
Keplerian limit but correspond to different initial conditions.
For the  ``\red{red}'' series the initial conditions correspond to the ``North Pole'' at the top of the Keplerian circle,
and for the ``\blue{blue}'' series they correspond to the ``Far-East'' one. The ``\purple{purple}'' series has a ``South-West'' initial condition.
Varying the strength of the perturbation from weak ($\omega_\rho=0.1$) through intermediate ($\omega_\rho=1$) 
to strong ($\omega_\rho=10$) deforms
the trajectory from the \cyan{pure Keplerian circle} (\cyan{\bf dashed cyan}) to the \magenta{pure anisotropic oscillator} (\magenta{\bf dotted magenta}). 
}
\label{2DKeplerOsc}
\end{figure}

Further insight is gained by studying some classical trajectories. Our strategy is to start with the planar Kepler problem and then consider what happens when
the relative strength of the perturbing  oscillator, represented by $\omega_\rho$, is varied from weak to strong.
The three rows of Figs. \ref{2DKeplerOsc}, \ref{2DQD} 
 correspond to identical initial conditions, namely to
\beqa\left\{
\barraynb{llllll}
x(0)=0,\quad &z(0)=1,
\quad
&\dot{x}(0)=1,
\quad
&\dot{z}(0)=0,
&\qquad
&\hbox{\red{red}}
\\
x(0)=1,\quad &z(0)=0,
\quad
&\dot{x}(0)=0,
\quad
&\dot{z}(0)=1
&\qquad
&\hbox{\blue{blue}}
\\
x(0)=-\frac{1}{2},\quad 
&z(0)=-\frac{1}{2},
\quad
&\dot{x}(0)=-\frac{1}{2},
\quad
&\dot{z}(0)=\frac{1}{2}
&\qquad
&\hbox{\purple{purple}}
\earraynb\right.\;,
\label{2dincond}
\eeqa
with the pure Keplerian and oscillator cases indicated in \cyan{dashed cyan} and \magenta{dotted magenta}, respectively.

The same conventions is used later below for their 3D extensions in Figs. \ref{3DKeplerOsc} and \ref{3DQD}, where we start from a point on the 2D trajectory, but we add some non-trivial $y$-initial condition.

\subsection{Attractive case $a>0$}\label{AttrPlots}

We first consider the attractive Coulomb/Kepler interaction, $a>0$.
As a result of the anisotropy [$\tau=2$] of the oscillator, the trajectories
 show a strong dependence on the initial conditions. Due to the complexity of the problem, we  limit our investigations therefore to the particular case 
\beq
|\vx(0)|=1
\qquad
|\dot{\vx}(0)|=1,
\label{CircInitCond}
\eeq
with the oscillator strength $\omega_\rho$
sweeping from small to big value,
Hence $a=1$ and the initial Keplerian trajectory is the unit circle \footnote{In the proof of Bertrand's Theorem \cite{Arnold}, which says that the only spherically symmetric potentials all of whose trajectories are closed, are the Kepler problem and the isotropic oscillator, one also starts with  circular motions and then asks which perturbations do yield closed trajectories.
}.
Turning on the anisotropic oscillator manifestly squeezes the initial circle. For 
$\omega_\rho\to\infty$ the trajectories converge to those of pure  $2:1$ anisotropic oscillator, indicated in \magenta{dotted magenta}. 

\subsection{The repulsive case $a<0$}\label{ReprPlots}

The Coulomb interaction between the  electrons which constitute genuine Quantum Dots is \emph{repulsive}, though~: $a\propto -e^2 < 0$. In the  pure Coulomb case, all trajectories are unbounded, namely hyperbolas. Switching on the harmonic trap converts the latter into bound ones, however. Intuitively, farther one goes stronger the harmonic force becomes, and ultimately wins against the weakening Coulomb repulsion.
The only effect is that the Dot becomes somewhat larger.

\begin{figure}
\null\hskip-10mm
\includegraphics[scale=.45]{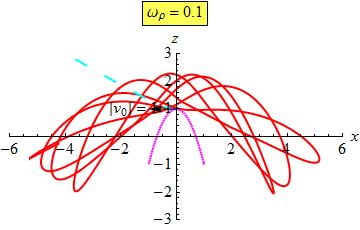}
\includegraphics[scale=.45]{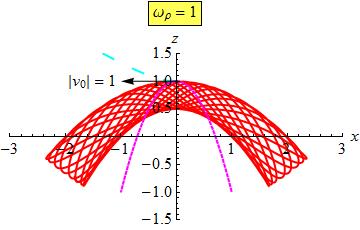}
\includegraphics[scale=.45]{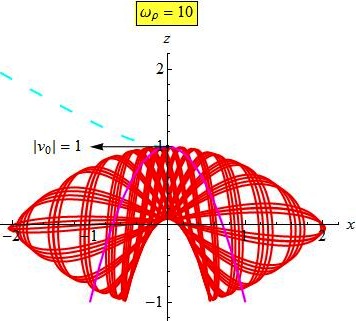}
\\
\null\hskip-10mm
\includegraphics[scale=.44]{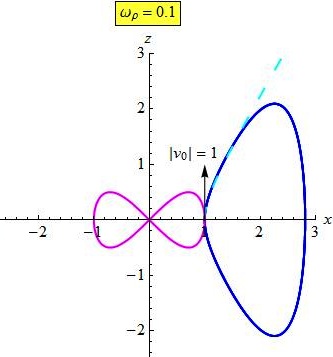}
\includegraphics[scale=.44]{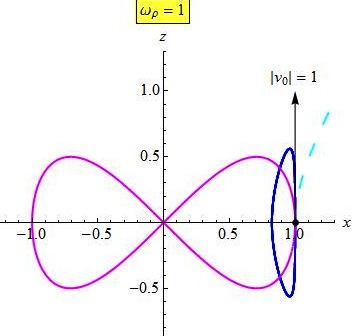}
\includegraphics[scale=.44]{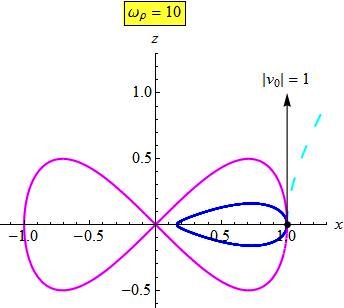}
\\
\null\hskip-14mm
\includegraphics[scale=.45]{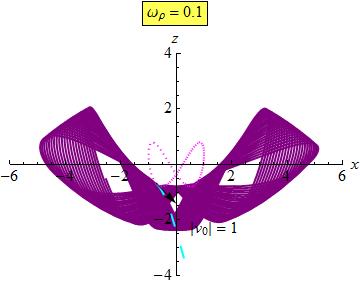}
\includegraphics[scale=.45]{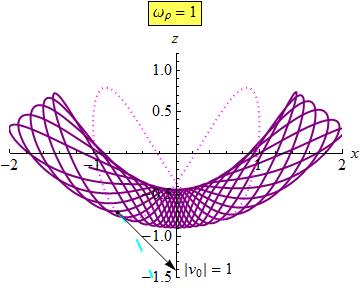}
\includegraphics[scale=.45]{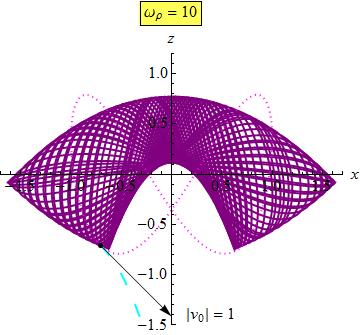}
\vspace{-5mm}
\caption{\it Trajectories for a repulsive Coulomb potential in the plane, perturbed with a $\tau=\omega_z/\omega_\rho=2$ anisotropic oscillator. 
 Turning on the harmonic oscillator from weak ($\omega_\rho=0.1$) through intermediate ($\omega_\rho=1$) 
to strong ($\omega_\rho=10$) deforms the initial \cyan{Keplerian hyperbola} (in \cyan{dashed cyan}) into closed ``potato" and ultimately into the  \magenta{dotted magenta} ``horizontal $8$" of the \magenta{pure oscillator}.
All intial conditions are tangent to the Keplerian hyperbola, but in various positions. The ``\blue{blue}'' series corresponds to the ``Near-East'' and
the  ``\red{red}'' series  corresponds to the bottom of the Keplerian hyperbola.
 The ``\purple{purple}'' series has a ``South-West'' initial condition.
}
\label{2DQD}
\end{figure}

A couple of trajectories are shown on Fig. \ref{2DQD}.
Here, all motions start from a point on the \cyan{Keplerian hyperbola} 
(in \cyan{dashed cyan})
  with identical initial conditions  as in the attractive case in Fig. \ref{2DKeplerOsc}
[as the colors suggest].
For $\omega_\rho\to\infty$ the trajectories tend to those of the pure \magenta{anisotropic oscillator} (in \magenta{dotted magenta}).

\subsection{Return to 3D}\label{3DPlots}

\begin{figure}
\null\hskip-14pt
\includegraphics[scale=.41]{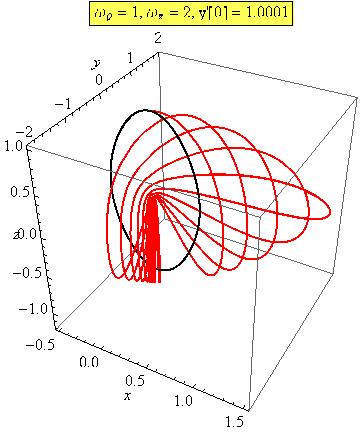}
\includegraphics[scale=.41]{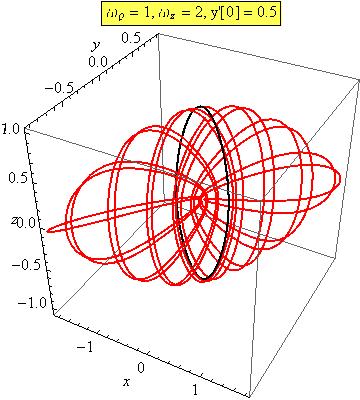}
\includegraphics[scale=.41]{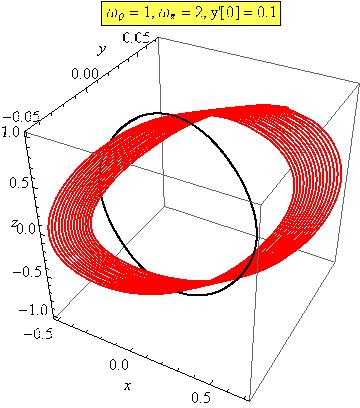}
\\
\null\hskip-14mm
\includegraphics[scale=.44]{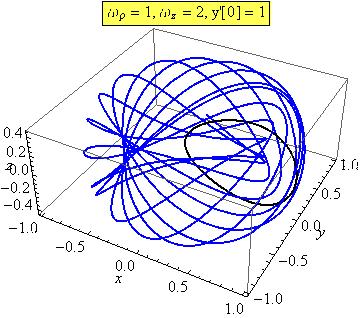}
\includegraphics[scale=.44]{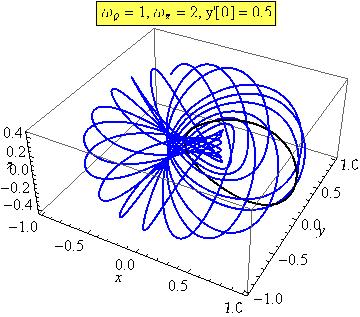}
\includegraphics[scale=.44]{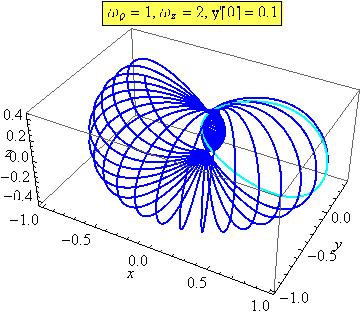}
\\
\null\hskip-14mm
\includegraphics[scale=.42]{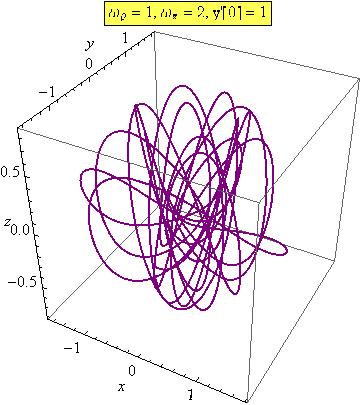}
\includegraphics[scale=.42]{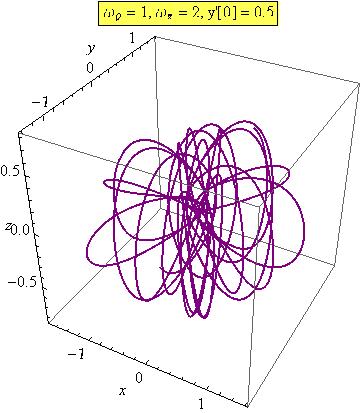}
\includegraphics[scale=.42]{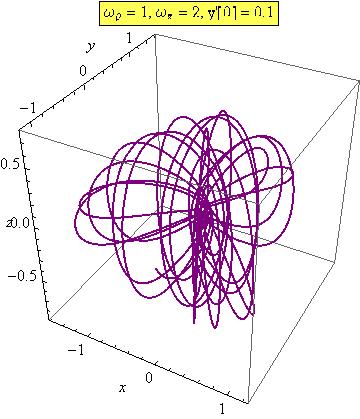}
\vspace{-3mm}
\caption{\it Some   3D trajectories in the perturbed Kepler problem with anisotropy
$\tau=2$. As suggested by using the same colors,
all figures have initial conditions lying on the 2D orbits of Fig. \ref{2DKeplerOsc} with intermediate coupling $\omega_\rho=1$, but with non-vanishing initial $y$-velocities $\dot{y}(0)=1,\,0.5,\,0.1$.   
The initial 2D trajectories in the $x-z$ plane are indicated in {\bf black}.
}
\label{3DKeplerOsc}
\end{figure}

\begin{figure}
\null\hskip-12mm
\includegraphics[scale=.45]{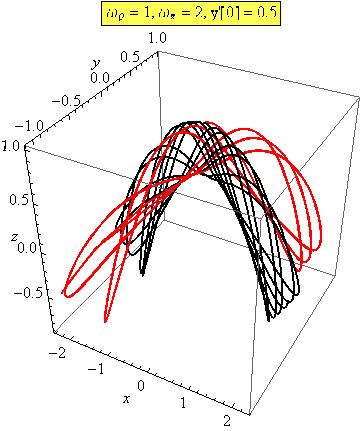}
\includegraphics[scale=.45]{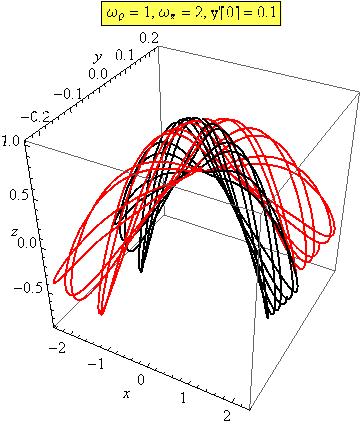}
\includegraphics[scale=.45]{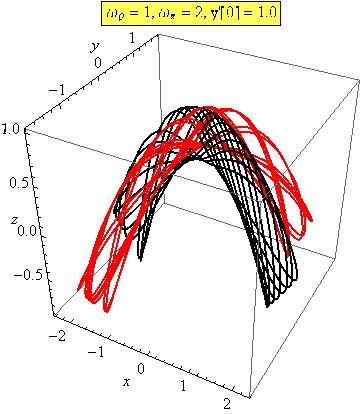}
\\
\null\hskip-14mm
\includegraphics[scale=.46]{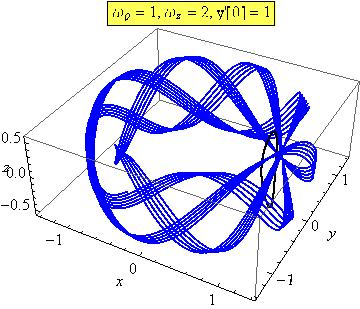}
\includegraphics[scale=.46]{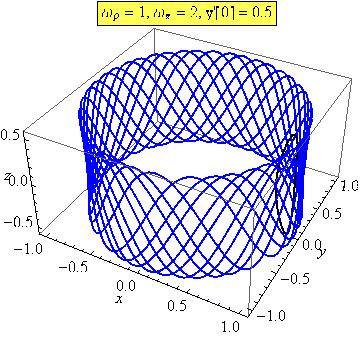}
\includegraphics[scale=.46]{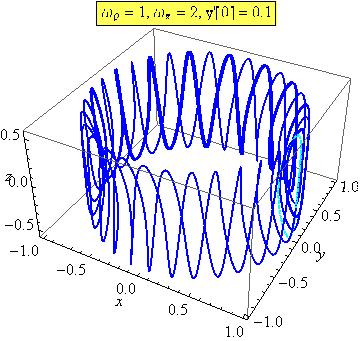}
\\
\null\hskip-14mm
\includegraphics[scale=.46]{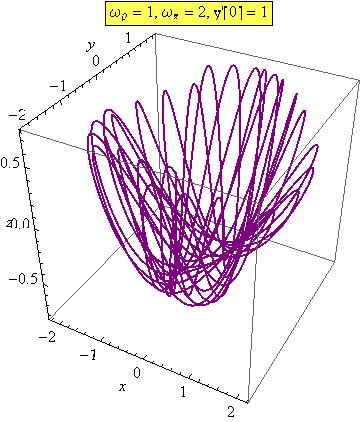}
\includegraphics[scale=.46]{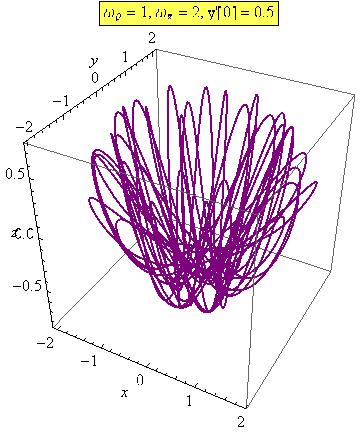}
\includegraphics[scale=.46]{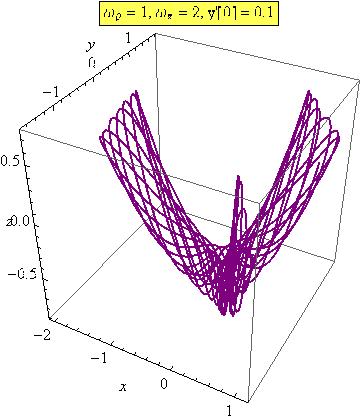}
\vspace{-6mm}
\caption{\it Some 3D trajectories in the repulsive Coulomb perturbed by a $\tau=2$ anisotropic  oscillator.  
The initial 2D trajectories (in {\bf black}) in the $x-z$ plane are only shown for the \red{red} series.
}
\label{3DQD}
\end{figure}

Relaxing the constraint $L_z\equiv p_\varphi=0$ in (\ref{Lzconstraint}) plainly allows us to recover our  3D description.
For the coordinates $(\xi,\eta,\varphi)$   
separability guaranteed when $\tau=2$. 
(\ref{QDRLpar}) generalizes the
planar conserved quantity $K_z$ in (\ref{QDRLphi0}).
Some  trajectories
 are shown on Fig. \ref{3DKeplerOsc},  \footnote{For $\dot{y}(0)$=1 the ``red'' solution on Fig. \ref{3DKeplerOsc} develops a strange singularity whose origin is unclear for us as yet.}
 allowing us to
check the conservation of $K_z$ also 
numerically.

\subsection{The curious $1/2$ case}\label{tauhalf}

It follows from our general theory that, in 3D, the values $\tau=1$ and $2$ are the \emph{only separable cases}. Simonovi\'c et al. \cite{Simon} observe, however, that, \emph{for $L_z=0$  states},
 the system is \emph{integrable} also for $\tau=1/2$. See also \cite{Alhassid,Blumel}.

Let us explain how this comes about. 
[We again turn to classical mechanics].
Consider the Kepler+axially symmetric oscillator Hamiltonian in (\ref{KeposciHam}), and introduce  new, ``twisted'' variables by rotating by $45$ degrees in $\xi-\eta$ space,
\beq
\mu=\frac{\xi+\eta}{\sqrt2},
\qquad
\nu=\frac{\xi-\eta}{\sqrt2},
\label{muvu}
\eeq
completed with $\varphi$.
Remarkably, 
\beq\barraynb{lll}
\xi\eta=\displaystyle\frac{\mu^2-\nu^2}{2}=\rho,
&\qquad
&\displaystyle\frac{\xi^2-\eta^2}{2}=\mu\nu=z,
\\[6pt]
\xi^2+\eta^2=\mu^2+\nu^2=2r,
&\qquad\qquad
&p_\xi^2+p_\eta^2=p_\mu^2+p_\nu^2,
\earraynb
\eeq
i.e., the coordinate transformation $(\xi,\eta)\to (\mu,\nu)$ 
interchanges $\rho$ and $z$ while leaving $r$  and $\vp^2$ invariant.
Then it follows that, expressed in terms of the new coordinates $\mu$ and $\nu$,
 \emph{$H_{Kepler}$} will have the same form as  (\ref{KeposciHam})
 \emph{with the exception of the  $p_\varphi^2$-term}. The latter changes as
\beq
\frac{p_\varphi^2}{\rho^2}=
\frac{p_\varphi^2}{\xi^2\eta^2}\quad\to\quad
\frac{4p_\varphi^2}{(\mu^2-\nu^2)^2}\,.
\eeq
The equation is hence form-invariant  only when this term is switched off by putting 
\beq
L_z\equiv p_\varphi=0,
\label{Lzconstraintbis}
\eeq
cf. (\ref{Lzconstraint}).
In other words, 
interchanging $\rho$ and $z$ is not a symmetry of the full 3-metric
$d\rho^2+dz^2+\rho^2d\phi^2$
written in cylindrical coordinates,
and hence not a symmetry of the  full kinetic term in the free Hamiltonian
$
\half\big(p_z^2 + p_\rho^2+{p_\phi^2}/{\rho^2}\big)
$
unless $p_\phi=0$.
Moreover, the exchange of $\rho$ and $z$ is not a global symmetry because $z$ ranges over all the reals while
$\rho$ ranges only over the positive reals.

The oscillator potential $V_{osc}$ transforms in turn as
\beq
\frac12\left[\omega_\rho^2\,\xi^2\eta^2+
\big(\frac{\omega_z}{2}\big)^2(\xi^2-\eta^2)^2\right]
\;\to\;
\frac12\left[\big(\frac{\omega_\rho}{2}\big)^{2}\,\big(\mu^2-\nu^2\big)^2+
\omega_z^2\,\mu^2\nu^2\right]\,,
\eeq
which are of the same form as written with $\xi$ and $\eta$, 
\emph{up to interchanging
the planar and vertical frequencies},
\beq
\omega_\rho\;\Longleftrightarrow\;\omega_z.
\label{freqinter}
\eeq
Hence, it is now the 
\beq
\tau=\frac{\omega_\rho}{\omega_z}
=\frac{1}{2}
\eeq 
case which is separable in the new coordinates
--- but  only \emph{when the constraint (\ref{Lzconstraint}) holds also}.
 
We note that the $(\mu,\nu)$ in (\ref{muvu}) can also be  considered as coordinates
in our vertical $(x-z)$ plane, 
\beq
x =\frac{1}{2}(\mu^{2}-\nu^{2}),
\qquad
z=z_+=\mu^{2}\nu^{2},
\label{twistcoord}
\eeq
This coordinate system suffers however of the same problems as $(\xi,\eta)$ in (\ref{2Dparabcoord}): while now
$-\infty <x < \infty$ we necessarily have $z=z_+ >0$ so that only the upper half-plane is covered, and (\ref{twistcoord}) has to be supplemented with $z=z_-=-\mu^{2}\nu^{2}<0$.

Having  understood these subtleties, 
 $(\xi,\eta)\to (\mu,\nu)$ amounts of rotating the plane by $90^{\smallcirc}$, $(x,z)\to(z,-x)$.
In terms of (\ref{twistcoord}), the Kepler+oscillator system is
precisely (\ref{KeposciHam}) with the $p_\varphi$-term switched off and
the frequencies interchanged as in (\ref{freqinter}).
Our entire machinery can now be applied once over again, simply by trading $(\xi,\eta)$ for $(\mu,\nu)$. Separability is now obtained for 
\beq
\tau=\frac12.
\eeq
The first line from the conserved quantities (\ref{first_int}) is the Hamiltonian (\ref{xietaKOham}), up to changing the variables into $(\mu,\nu)$ and replacing 
$\omega_z$ with $\omega_\rho$.
The second line yields in turn
\beq
\displaystyle\frac1{2(\mu^2+\nu^2)}\left[
-\nu^2p_\mu^2+\mu^2p_\nu^2
 \right]-a\frac{\mu^2-\nu^2}{\mu^2+\nu^2}
-
\frac{\omega_\rho^2}{4}
\underbrace{(\mu^2\nu^2)(\frac{\mu^2-\nu^2}{2})}_{z^2\rho}
\label{QDRLphi0x}
\eeq
which is also the same as $K_z^0$ in (\ref{QDRLphi0}) after the
interchange $(\xi,\eta)\leftrightarrow(\mu,\nu)$, as
 expected.
Moreover, using 
$
{p_\mu^2-p_\nu^2}=2p_\xi{}p_\eta
$
(\ref{QDRLphi0x}) reduces, for $\omega_\rho=0$, to   $-\big(-E/2\big)^{1/2}\,K_x$ in (\ref{2DKx}).

Note that  the correction term in (\ref{QDRLphi0x}) which arises due to the $\tau=1/2$ oscillator is now
$ 
-(\omega_\rho/2)^2\,z^2\rho,
$ 
as expected from the interchange 
$\rho\leftrightarrow z$, cf. (\ref{QDRLphi0}).

Turning off the anisotropic oscillator restores the
rotational and indeed the full $\Orth(3)$ symmetry, with the two components of the planar Runge-Lenz corresponding to separability in the two respective coordinate systems.

\begin{figure}
\begin{center}
\includegraphics[scale=.45]{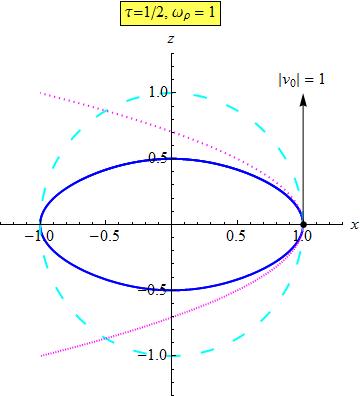}\qquad\qquad
\includegraphics[scale=.43]{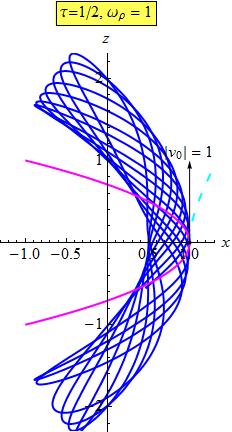}
\vspace{-8mm}
\end{center}
\caption{\it In the plane,  the (i) [attrative] Kepler and the (ii)  [repulsive] QD problem, perturbed by a 
$\tau=\omega_z/\omega_\rho=1/2$ oscillator is plainly separable in the twisted coordinates $(\mu,\nu)$, since the latter correspond to a rotation by $90^{\smallcirc}$ degrees,  interchanging the ``long'' and ``short'' directions and carrying $K_z$ into $-K_x$.
The  $\tau=1/2$-figure  is indeed the rotated
$\tau=2$-figure, in Figs. \ref{2DKeplerOsc}
and \ref{2DQD}, respectively.}
\label{2dKeplerOsc12}
\end{figure}
The regularity of the trajectories obtained for $\tau=1/2$ hints at an additional conserved quantity. So far, we derived such quantities from separability using the St\"ackel approach. Separability is, however,  not a
\emph{necessary}, only a \emph{sufficent}  condition for such a quantity, and we can, following Bl\"umel et al. \cite{Blumel}, proceed directly to search such a quantity.
Their strategy is to observe, firstly, that the  usual Keplerian Runge-Lenz vector is not conserved, 
\beq
\dot{\bK}_{Kepler}\neq0
\qquad\hbox{for}\qquad
\bK_{Kepler}=\bp\times\bL\,-\,a\frac{\br}{r}\,.
\label{KRL}
\eeq
If, however, $\dot{\bK}_{Kepler}$ happens to be a total time derivative,
$\dot{\bK}_{Kepler}=\dot{d\vC},
$
then 
\beq
\bK=\bK_{Kepler}-\vC
\eeq
will be conserved.

Let us first put $L_z=0$.
For the combined Kepler + axisymmetric oscillator our condition requires,
for the components written in cylindrical coordinates,
\beq\left\{\barraynb{cll}
\omega_\rho^2\left(\displaystyle\frac{\tau^2-2}{2}z\dot{(\rho^2)}+\dot{z}\rho^2\right)
&=&\dot{C}_z,
\\[10pt]
\omega_\rho^2\left(\displaystyle\frac{1-2\tau^2}{2}\rho\dot{(z^2)}+\tau^2\dot{\rho}\,z^2\right)
&=&\dot{C}_\rho,
\earraynb\right.
\label{Cdot}
\eeq
obtained by calculting $\dot{K}^{Kepler}$ using the eqns of motion,
\beq\left\{\barraynb{lllll}
\ddot{\rho}-\displaystyle\frac{L_z^2}{\rho^3}&=&
-\omega_\rho^2\rho&-&a\displaystyle\frac{\rho}{(\rho^2+z^2)^{3/2}},
\\[8pt]
\ddot{z}&=&-\tau^2\omega_\rho^2z&-&
a\displaystyle\frac{z}{(\rho^2+z^2)^{3/2}}.
\earraynb\right.
\label{Blumeleqmot}
\eeq

The conditions (\ref{Cdot}) 
require
\beq
\tau=2 \;\Rightarrow\;   C_z=\omega_\rho^2\rho^2z
\qquad\hbox{or}\qquad
\tau=\frac12 \;\Rightarrow\;  C_\rho=\smallover1/4\omega_\rho^2\rho z^2,
\eeq
which can not hold simultaneously, but provide us with  either of our two previous cases,
\beq\left\{\barraynb{cllll}
K_z^0&=&z\dot{\rho}^2-\dot{z}\rho\dot{\rho}
-a\displaystyle\frac{z}{\sqrt{\rho^2+z^2}}-\omega_\rho^2\rho^2z
&&\hbox{for}\quad\tau=2,
\\[16pt]
K_\rho^0&=&\rho\dot{z}^2-\dot{\rho}z\dot{z}
-a\displaystyle\frac{\rho}{\sqrt{\rho^2+z^2}}-\smallover1/4\omega_\rho^2\rho z^2
&&\hbox{for}\quad\tau=1/2
\earraynb\right.\quad.
\label{2DRLzrho}
\eeq

Restoring 3D by lifting the constraint $L_z=0$ merely requires,
in the separable case $\tau=2$,
a further correction term,
\beq
K_z=K_z^0+\frac{z}{\rho^2}L_z^2\,,
\eeq
which is indeed (\ref{QDRLpar}). 
 
In the integrable but non-separable case $\tau=1/2$ Bl\"umel et al. \cite{Blumel} find the \emph{quartic} conserved quantity
\beq
K^{(4)}=\big(K_\rho^0+\frac{L_z^2}{\rho}\big)^2+\big(K_\varphi^0\big)^2
+\omega_\rho^2(\rho^2+z^2)L_z^2\,,
\label{half3DRL}
\eeq
where
$K_\rho^0$ is the one in (\ref{2DRLzrho}), and
\beq
K_\varphi^0=K_\varphi^{Kepler}=
-\frac{\rho\dot{\rho}+z\dot{z}}{\rho}L_z\;.
\eeq
$K^{(4)}$ is hence the [squared] length of the planar expression in (\ref{2DRLzrho}),
corrected with terms which involve $L_z\neq0$.
For $L_z=0$ (\ref{half3DRL}) reduces to $K_x^2$,
the square of $K_x$ in (\ref{2DKx}) and/or
in (\ref{QDRLphi0x}). 
The conservation of (\ref{half3DRL}) can be checked directly using the equations of motion.

It is now easy to understand the fundamental difference between the two  semi-parabolic coordinates systems.
The standard one we denoted by $(\xi,\eta)$ are naturally extended from 2D to 3D by adding $\varphi$, which unifies the two local 2D-charts associated with $x_+$ and $x_-$, since $\cos\pi=-1$
produces exactly the desired sign change.

For the ``twisted coordinates $(\mu,\nu)$, however, the trick does not work: adding the polar angle $\varphi$ 
 does \emph{not}
change $z>0$ into $z<0$, and so half of the space still remains uncovered. 

\begin{figure}
\begin{center}\null\hskip-10mm
\includegraphics[scale=.41]{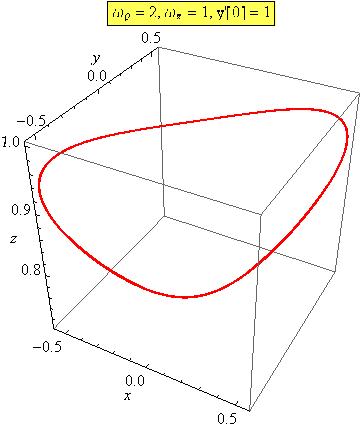}\quad
\includegraphics[scale=.48]{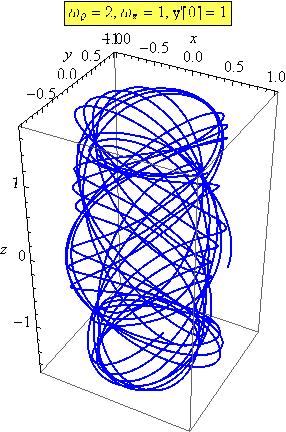}\quad
\includegraphics[scale=.45]{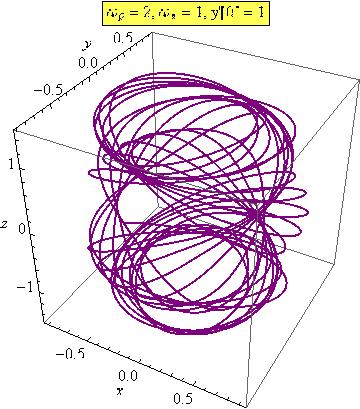}
\vspace{-8mm}
\end{center}
\caption{\it Trajectories in the integrable but non-separable case $\tau=\omega_z/\omega_\rho=1/2$
for various initial conditions.
}
\label{3dtauhalf}
\end{figure}

\section{The Quantum Picture}\label{Quantum}
 
Let us now outline, for completeness, how things behave at the quantum level, cf. Refs.  \cite{Simon, Simon2, Alhassid}. As it follows from our general theory, the
only separable coordinate systems are the spherical one, for $\tau=\omega_z/\omega_\rho=1$, and the semi-parabolic one, for $\tau=\omega_z/\omega_\rho=2$. 
The first one of these is routine-like, and below we only study therefore the second case. The Schr\"odinger equation (\ref{pertKpb}) for relative motion reads, 
\begin{equation}
\left[i\partial_t+\underbrace{\,\frac 1{2}\bigtriangleup+\frac{a}{r}\,}_{Kepler}-V_{osc}\right]\psi=0.
\label{tau2Kosc}
\end{equation}
In semi-parabolic coordinates (\ref{sparabc}), the Laplacian is
\beq
\bigtriangleup=\frac{1}{\xi^2+\eta^2}\left[
\frac{1}{\xi}\p_\xi\big(\xi\p_\xi\big)+
\frac{1}{\eta}\p_\eta\big(\eta\p_\eta\big)+
\left(\frac{1}{\xi^2}+\frac{1}{\eta^2}\right)
\right].
\label{sparLaplace}
\eeq
Our task is hence to solve,
\beq
\left[
2(\xi^2+\eta^2)i\partial_t+\frac{1}{\xi}\p_\xi\big(\xi\p_\xi\big)+
\frac{1}{\eta}\p_\eta\big(\eta\p_\eta\big)+
\Big(\frac{1}{\xi^2}+\frac{1}{\eta^2}\Big)\p_\varphi^2
+2a
-\omega_\rho^2\big(\xi^6+\eta^6\big)\right]\psi=0.
\label{tau2QKO}
\eeq
Then, consistently with the Robertson theorem \cite{Cordani}, for $\tau=2$ i.e. for $\omega_z=2\omega_\rho$
the Ansatz 
\beq
\psi(\xi,\eta,\varphi,t)=({\xi\eta})^{-1/2}u(\xi)v(\eta)e^{im\varphi}e^{-iEt}
\label{parsepAnsatz}
\eeq
 separates the Schr\"odinger equation. Putting $\xi_1=\xi$ and $\xi_2=\eta$, we have,
\beqa
\left[
\frac{\,\,d^2}{d\xi_i^2}+2E\xi_i^2
-\frac{m^2-\smallover1/4}{\xi_i^2}+A_i-\omega_\rho^2\,\xi_i^6\right]u(\xi_i)&=&0,
\qquad
i=1,2,
\label{Schxi}
\eeqa
where the separation constants must satisfy the constraint
\beq
A_1+A_2=2a.
\label{Kepsepconst}
\eeq
Note that (\ref{Kepsepconst}) is indeed the only trace of the Kepler term. For a pure oscillator, $a=0$.

We now study (\ref{Schxi}) dropping the subscript $i=1,2$. Firstly, the $\xi^{-2}$ term can be eliminated, just like for Kepler, but putting 
$u=\xi^{(|m|+1/2)}U$, yielding,
\beqa
\left[
\frac{\,\,d^2}{d\xi^2}+2E\xi^2
+A-\omega_\rho^2\,\xi^6\right]U(\xi)=0.
\label{SchxiU}
\eeqa
Regularity of $\psi$ at the origin is then guaranteed if $U$ and $V$ remain finite near the origin, 
\beq
\psi(\xi,\eta,\varphi,t)\approx \rho^{|m|}U(\xi)V(\eta)e^{im\varphi}e^{-iEt}
\qquad
\xi,\eta\approx0,
\label{smallrho}
\eeq
where we used $\xi\eta=\rho$. 
For large $\xi$  instead, the $6^{th}$-order oscillator term  dominates. Dropping all other terms yields
$
{d^2U}/{d\xi^2}
-\omega_\rho^2\,\xi^6\,U\;\approx\;0,
$
whose approximate solution which vanishes at infinity is $U(\xi)\approx
e^{-|\omega_\rho|\,\xi^4/4}$. For large 
$\xi$ and $\eta$ we have, hence, essentially a pure oscillator,
\beq
U(\xi,\eta)\approx
e^{-|\omega_\rho|\,(\xi^4+\eta^4)/4}
=e^{-|\omega_\rho|\,(\rho^2+2z^2)/2},
\qquad
\xi,\eta\to\infty.
\label{largeU}
\eeq

More generally, our Eqn. (\ref{SchxiU}) is, up to shifting the constraint (\ref{Kepsepconst}) from $0$ to arbitrary constant $a$, \emph{identical} to the one which describes the \emph{pure $2:1$ anisotropic oscillator in the plane} \cite{Boyer} \footnote{Alternatively, Eqn. (\ref{SchxiU}) describes a  1D \emph{anharmonic oscillator} with a $6th$-order  potential
$-\Omega^2\xi^2+\omega_\rho^2\xi^6$ \cite{Truong}. }.

For a detailed analytical study of Eqn. (\ref{SchxiU}) the Reader is referred to the literature, and to Refs.
\cite{Blumel,Boyer,Truong} in particular. Some numerical solutions are plotted  below.


We now turn to solving 
 Eqns. (\ref{Kepsepconst})-(\ref{SchxiU})
 numerically for bound states. 
Let us observe that it is a 
\emph{two-parameter} problem: the equation to be solved involves both the separation constant $A$ and the energy, $E$, which should be correlated.
 
For pure Kepler, or for the isotropic oscillator, the two separation constants can be unified into one. Then one can find the single ``good'' value which makes the solution bounded
either analytically (namely from the poles of the hypergeometric function \cite{LandauLifshitz}), or also numerically.


Reduction to a one-parameter problem similar procedure would also work for the 2D pure oscillator with frequencies $\omega_1$ and $\omega_2$ in Cartesian coordinates, when can proceed as follows. The natural product Ansatz splits the 
Schr\"odinger equation into two 1D problems, 
\beq
u_i''+[2\epsilon_i-\omega_i^2x_i^2]u_i=0,
\quad
\epsilon_1=\half(E-C),
\;
\epsilon_2=\half(E+C)
\quad\Rightarrow\quad
E=\epsilon_1+\epsilon_2.
\label{Car2DO}
\eeq
The two eqns have identical [namely 1D oscillator] form, and are coupled through $E$ and $C$. But the two constants are, however, unified into single ones.
Solving each of them independently for bound states  yields the possible ``good'' values of the energies, 
namely $\epsilon_i=\omega_i(n_i+\half)$. 
Then from (\ref{Car2DO}) we infer the 2D spectrum, 
\beq
E\equiv E_{n_1,n_2}=\epsilon_1+\epsilon_2=\omega_1(n_1+\half)+\omega_2(n_2+\half).
\label{Ospectrum}
\eeq
 For our $2:1$  system, in particular, $\omega_1=2\omega_2\equiv2\omega$, and
the 2D energy becomes one with a single principal quantum number $N$,
\beq
E=E_N=\omega\,(N+\smallover3/2),
\qquad
N=2n_1+n_2.
\label{21Ospect}
\eeq
The energy levels are therefore $[N/2]+1$-times degenerate, as it follows from the formula for $N$. Keeping $N$ fixed also tells us which individual solutions should be paired together.

To solve the problem in parabolic coordinates, we would need a relation between $E$ and $A$ similar to the one above that we don't have, though, let alone for the pure oscillator \footnote{In the pure oscillator case, we can do the following trick. We just \emph{know} from the
Cartesian result the energy spectrum, so we simply eliminate the parameter $E$ by putting its value (\ref{21Ospect}) into the equation to be solved. This leaves us with one
separation constant alone, $A$, en we know from (\ref{Kepsepconst}) that the two solutions with separation constants $A$ and $-A$ should be paired. (It is known, moreover, that if $u(\xi)$ works for $A$, then $u(-\xi)$ will work for $-A$, and is hence suitable for the pure oscillator). 
Having fixed the energy $E=E_N$, the computer provides us with
 a collection of good separation constants $A_k$, $k=1,\dots, [N/2]+1$ which provide us with all bound states with the same energy. }.

\begin{figure}
\includegraphics[scale=.34]{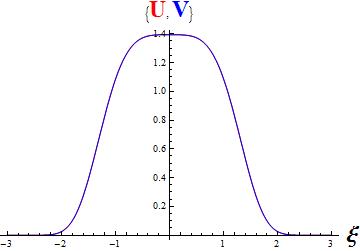}
\\[6pt]
\includegraphics[scale=.34]{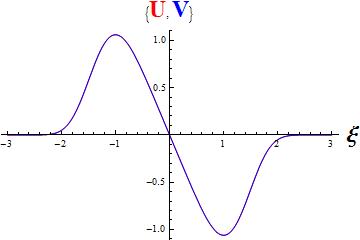}
\\[6pt]
\includegraphics[scale=.34]{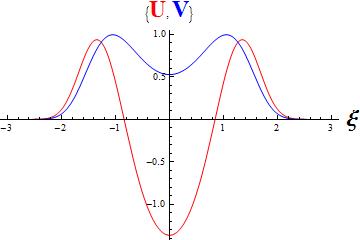}
\includegraphics[scale=.34]{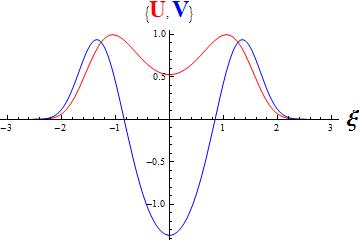}
\\[6pt]
\includegraphics[scale=.34]{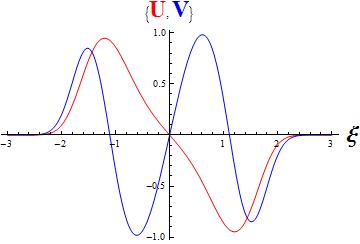}
\includegraphics[scale=.34]{PureOscN3a.jpg}
\\[6pt]
\includegraphics[scale=.34]{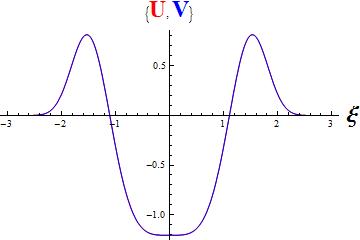}
\includegraphics[scale=.34]{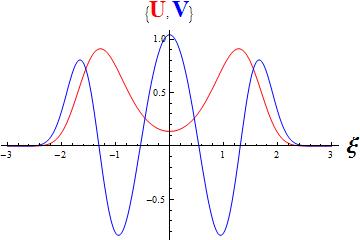}
\includegraphics[scale=.34]{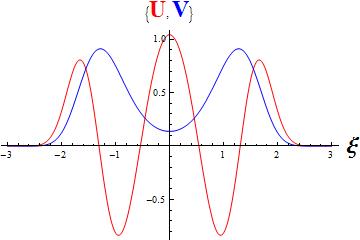}
\\[6pt]
\includegraphics[scale=.34]{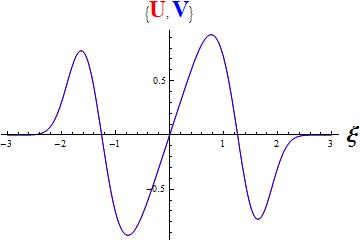}
\includegraphics[scale=.34]{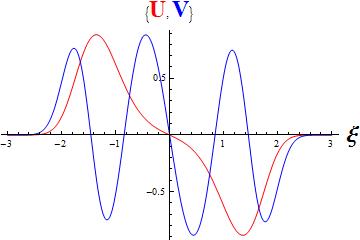}
\includegraphics[scale=.34]{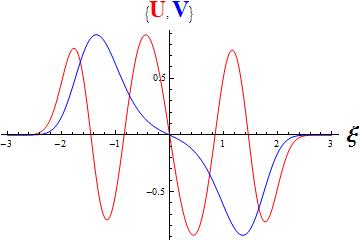}
\\
\vspace{-2mm}
\caption{\it Wave functions for the pure 2:1 oscillator ($\omega=1$) in the plane  for principal quantum numbers
$N=0,1,\dots,5$. $\red{U}$ is  plotted with \red{red} 
and its pair $\blue{V}$ in \blue{blue}.
}
\label{QOfig}
\end{figure}

So far for the oscillator alone.
But in the coupled oscillator + Kepler case, the problem is plainly \emph{not separable} in Cartesian coordinates, and so we can not determine the exact energy spectrum separately, and a two-parameter search for bound states had to be developed, providing us with Fig. \ref{QOfig} and Table \ref{pureOtable}, as well as with Figs. 
9, 10
 and  Table \ref{OKtable}, respectively.

Fig. \ref{QOfig} shows the solutions obtained for the pure $2:1$ oscillator. The energy values and degeneracies found numerically are consistent with the exact results. This search can be viewed, therefore, as a test for our two-parameter search.

The results listed in Table \ref{OKtable} and illustrated on Figs. 
9 and 10 
 show that turning on the Kepler interaction reduces the energy. This is clear from that
for the attractive  Kepler interaction 
$a>0$ (i) the energy is negative; moreover, (ii) The gravitational attraction it pulls closer the charges, reducing also the oscillator-energy.   
It is also interesting to observe (see Table \ref{OKtable} and Fig. 
 10
 that the Kepler term lifts the three-fold degeneracy of the $N=4$ pure-oscillator  states, splitting the triplet into a singlet plus two, doubly-degenerate states with slightly higher energy.
 
\begin{table}
\begin{tabular}{|c|c|c|c|}
\hline
Princ. quant. number
&Energy
&Separ. const.
&degeneracy
\\
\hline
$N=0$
&$E=\smallover3/2$
&$A=0$
&$d=1$
\\
\hline
$N=1$
&
$E=\smallover5/2$
&$A=0$
&$d=1$
\\
\hline
$N=2$
&$E=\smallover7/2$
&$A=\pm4.89898$
&$d=2$
\\
\hline
$N=3$
&
$E=\smallover9/2$
&$A=\pm4.89898Ê$
&$d=2$
\\
\hline
$N=4$
&
$E=\smallover{11}/2$
&$A=0,\pm8Ê$
&$d=3$
\\
\hline
$N=5$
&
$E=\smallover{13}/2$
&$A=0,\pm11.3137$
&$d=3$
\\
\hline
\end{tabular}
\caption{\it Numerical results for the pure 2:1 oscillator with $\omega=1$. For $N=2k+1$ odd the good values of the separation constants $A$ come in pairs of opposite signs, to which $A=0$ is added for $N=2k$ even.}
\label{pureOtable}
\end{table}
\medskip

\goodbreak
\begin{table}
\begin{tabular}{|c|c|c|}
\hline
Energy
&Separation const.
&degeneracy
\\
\hline
$E_0=0.228586$
&$A_1=1$
&$d=1$
\\
\hline
$E_1=2.00297$
&$A_1=1$
&$d=1$
\\
\hline
$E_2=2.91222$
&$A_1=-1.7712, A_1=3.7712$
&$d=2$
\\
\hline
$E_3=4.10518$
&$A_1=-3.81344,5.81344$
&$d=2$
\\
\hline
$E_4=4.78076$
&$A_1=1$
&$d=1$
\\
\hline
$\tilde{E}_{4}=5.13544$
&$A_1=-6.89357, 8.89357$
&$d=2$
\\
\hline
\end{tabular}
\caption{\it Numerical results for the coupled 2:1 oscillator + attractive Kepler potential. The separation constant $A_2$ is determined by the constraint (\ref{Kepsepconst}). We took $a=1$ and $\omega_\rho=1$.}
\label{OKtable}
\end{table}

\begin{figure} 
\includegraphics[scale=.34]{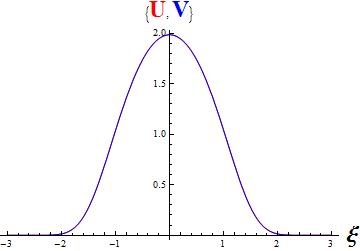}
\\[8pt]
\includegraphics[scale=.34]{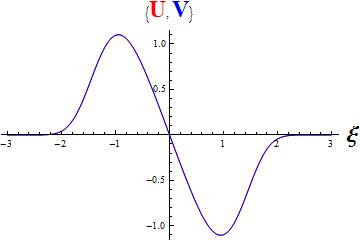}
\\[8pt]
\includegraphics[scale=.34]{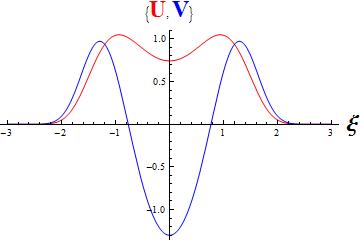}
\includegraphics[scale=.34]{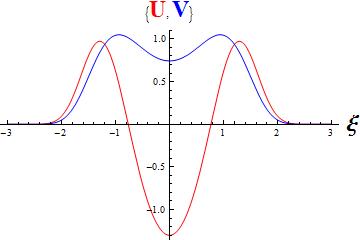}
\\[8pt]
\includegraphics[scale=.34]{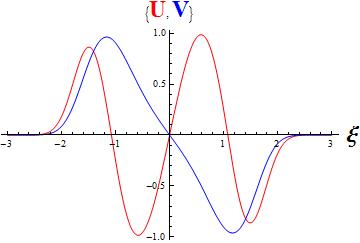}
\includegraphics[scale=.34]{KepOsca2N3a.jpg}
\label{QOKfig-1}
\caption{\it The lowest-energy wave functions of the coupled 2:1 oscillator perturbed with an attractive Kepler potential. 
$a=1$ and $\omega=1$.}
\end{figure}

\begin{figure} 
\includegraphics[scale=.34]{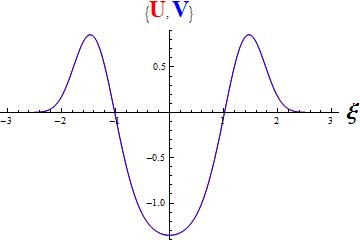}
\\[6pt] 
\includegraphics[scale=.34]{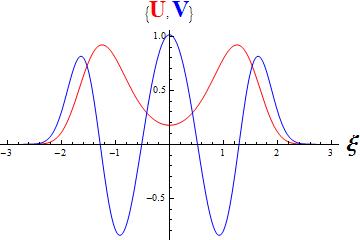}
\qquad\qquad
\includegraphics[scale=.34]{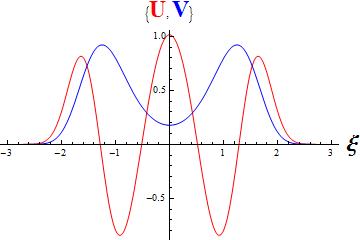}
\label{QOKfig-2}
\caption{\it The Kepler perturbation splits the $N=4$ triplet of states of the pure oscillator into a singlet plus a slightly higher-energy doublet, cf. Table \ref{OKtable}.}
\end{figure}

The combined case with repulsive (Coulomb-type) interaction is presented.
in Table \ref{ORKtable}
and on Fig. 11.

\begin{table} 
\begin{tabular}{|c|c|c|}
\hline
Energy
&Separ. const.
&degener
\\
\hline
$E_0= 2.38668$
&$A_1=-1$
&$d=1$
\\
\hline
$E_1=4.04956$
&$A_1=-4.01553, 2.01553$
&$d=2$
\\
\hline
$E_2=5.85676$
&$A_1=-9.14289, 7.1428$
&$d=2$
\\
\hline
\end{tabular}
\caption{\it Numerical results for the coupled oscillator + repulsive Coulomb potential, relevant for Quantum Dots.
$a=-1, \omega=1$.}
\label{ORKtable}
\end{table}

\begin{figure} 
\includegraphics[scale=.34]{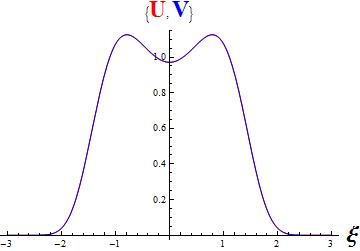}
\\[6pt]
\includegraphics[scale=.34]{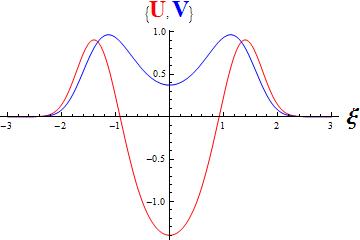}
\includegraphics[scale=.34]{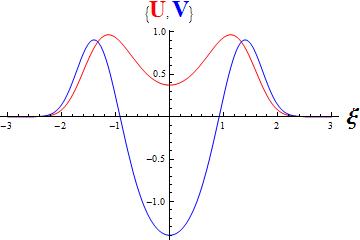}
\\[6pt]
\includegraphics[scale=.34]{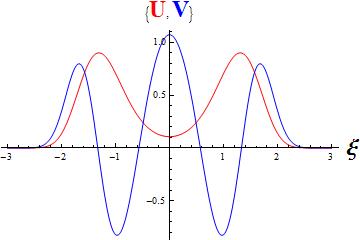}
\includegraphics[scale=.34]{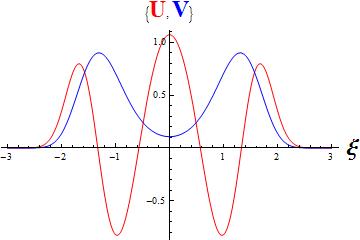}
\label{QORKfig}
\caption{\it The lowest-energy states for the 2:1 oscillator coupled to a repulsive Coulomb potential, relevant for Quantum Dots.}
\end{figure}

\section{Further separable perturbations}\label{seppot}

More generally, our  trick plainly works for any axial potential which satisfies, in parabolic coordinates, the separability condition (\ref{seppar}). 
For example~:

\bequ
\item
Let us consider, e.g., the \emph{Hartmann potential} used in quantum chemistry \cite{Hartmann,KiWi},
\begin{equation}
V^{Hartmann} =\frac{a}{\rho^2}=\frac{a}{\xi^2\eta^2}
\label{Hartmann}
\end{equation}
in (semi)parabolic coordinates. The separability condition (\ref{seppar}) is satisfied, since
\beq
(\xi^2+\eta^2)V=\frac{a}{\xi^2}+\frac{a}{\eta^2}
=f(\xi)+g(\eta).
\eeq
Eqn. (\ref{first_int}) provides us with three conserved quantities in involution. The generalized Runge-Lenz-type scalar $K_z$ is, in particular, of the form (\ref{QDRLpar}) and (\ref{QDRLord}),
respectively, but where the last, additional term is rather 
\beq
a\,\frac{\xi^2-\eta^2}{\xi^2\eta^2}=2a\frac{z}{\rho^2}.
\eeq

The system is separable also in spherical coordinates 
cf. \cite{Hartmann,KiWi}. The spherical St\"ackel quantities are 
(\ref{3DsphStaeckel}) except for the last contribution to 
$\underline{w} $ which, fixed by the potential,
should read now
\beq
\underline{w}=\barray{c}0\\
\frac{a}{\sin^2\theta}\\ 0
\earray .
\label{Hartmannw}
\eeq
The mutually commuting
conserved quantities are therefore $H,\, L_z^2/2-a$, and
the modified total angular momentum-square,
\beq
{\cal L}^2=\frac{L^2}{2}+\frac{a}{\sin^2\theta}\,,
\label{HartmannL2}
\eeq
as found before  \cite{KiWi}.

\item
Another example is provided by the \emph{constant perturbing
field} $\bE=E\hat{z}$ parallel to the magnetic field considered in the Stark effect \cite{Cordani}, 
\begin{equation}
V=Ez=\frac{E}{2}(\xi^2-\eta^2)
\quad\Rightarrow\quad
(\xi^2+\eta^2)V=\frac{E}{2}(\xi^4-\eta^4).
\label{Ezfield}
\end{equation}
The Runge-Lenz type scalar $K_z$ is proportional to the projection of the Runge-Lenz vector on the electric field, augmented with a correction term \cite{Redmond},
\beq
(\bL\times\bp-a\hat{\br})\cdot\bE-\frac12(\br\times\bE)^2.
\eeq
\item
General polynomial solutions to (\ref{seppar}) are obtained \cite{Komornik} for \emph{any integer} $n=0,\pm1,\dots$, $a=\const$, by
\beq
V_n=a\,\frac{\xi^{2n}+(-1)^{n+1}\eta^{2n}}{\xi^2+\eta^2}\,.
\label{npot}
\eeq
which is indeed manifestly separable. On the other hand, the algebraic identity
$$
\xi^{2n+2}+(-1)^{n+2}\eta^{2n+2}=
(\xi^{2}-\eta^{2})
(\xi^{2n}+(-1)^{n+1}\eta^{2n})-
(\xi^{2}\eta^{2})(\xi^{2n-2}+(-1)^{n}\eta^{2n-2})
$$
translates into
$
V_{n+1}=2z\,V_n-\rho^2\,V_{n-1},
$
proving by induction that $V_n$ is also axially symmetric. 
Similarly, the  identity
$$
\underbrace{\xi^{2n+2}+(-1)^{n+1}\eta^{2n+2}}_{\widetilde{V}_{n+1}}=
\underbrace{(\xi^{2}-\eta^{2})}_{2z}
\underbrace{(\xi^{2n}+(-1)^{n}\eta^{2n})}_{\widetilde{V}_n}+
\underbrace{(\xi^{2}\eta^{2})}_{\rho^2}
\underbrace{(\xi^{2n-2}+(-1)^{n-1}\eta^{2n-2})}_
{\widetilde{V}_{n-1}}
$$
shows that
\beq
\widetilde{V}_n
=a\,\frac{\xi^{2n}+(-1)^{n}\eta^{2n}}{\xi^2+\eta^2}
\label{ntildepot}
\eeq
is also separable and axially symmetric, providing us with a second doubly-infinite tower of
axially symmetric separable potentials.

For $n=-1$ we get \cite{Makar,KiWi}
\beq
\widetilde{V}=\frac{1}{2r}\left(\frac{1}{\xi^2}-
\frac{1}{\eta^2}\right)=-\frac{1}{r}\frac{z}{\rho^2}=-\frac{\cos\theta}{r^2\sin^2\theta}\,.
\label{makarovpot}
\eeq
Some further interesting cases are listed in Table \ref{tableau}.

\begin{table}
\begin{tabular}{|l|l|l|}
\hline
$n=0$
&$V=0$
&trivial
\\
\hline
$n=1$
&
$V=1$
&trivial
\\
\hline
$n=2$
&
$V=Ez$
&Stark effect
\\
\hline
$n=3$
&
$V=\rho^2+2z^2$
&1:2 oscillator
\\
\hline
$n=-1$
&
$V=\displaystyle\frac{1}{\rho^2}=\displaystyle\frac{1}{r^2\sin^2\theta}$
&Hartmann potential
\\[6pt]
\hline\hline
$n=0$
&$\widetilde{V}=\displaystyle\frac{1}{r}$
&Coulomb
\\[6pt]
\hline
$n=1$
&$\widetilde{V}=\displaystyle\frac{2z}{r}$
& ?
\\
\hline
$n=-1$
&
$\widetilde{V}=\displaystyle\frac{z}{r\rho^2}=
\displaystyle\frac{\cos\theta}{r^2\sin^2\theta}
$
& Makarov et al.\\[6pt]
\hline
\end{tabular}
\caption{\it Some potentials which are separable in parabolic coordinates.}
\label{tableau}
\end{table}
\eequ
 
Similar calculations show that, in the two remaining coordinate systems, no perturbing potential can be
added while preserving separability, though.

\section{Conclusion}

To explain the findings of Simonovi\'c et al. about the separability of quantum dots \cite{Simon} has been to trade first
the constant magnetic field for a pure axially symmetric oscillator by switching to rotating coordinates. 

The hydrogen atom is separable in four appropriate coordinate systems \cite{Cordani};
then we asked~: ``which potentials can be added so that separability is preserved in one of those coordinates~?"
The answer we found says that, apart of the  expected spherical
case, separability can be achieved  in
\emph{parabolic coordinates} for any axial potential which satisfies the separability condition (\ref{seppar}). 

For the  harmonic trap considered in the QD problem \cite{Simon} this requires a 2:1 anisotropy, cf. (\ref{2:1}).

To gain further insight, we found it convenient to first restrict the system to the vertical $x-z$ plane.
Then, removing the constraint $L_z=p_\varphi=0$, allowed us to recover the 3D motion and its properties. 

More general separable solutions, beyond the 2:1 oscillator, arise, though, some of them  listed in Table I \footnote{How could we find so many solutions~?  The intuitive answer is that the separability condition (\ref{seppar}) is not very restrictive. In the spherical case, which merely requires a radial potential.}.
These cases can plainly  be combined due to the additivity of both the functions $f(\xi)$ and $g(\eta)$ and of the potentials cf. (\ref{seppar}). One can, for example,
put the QD into an additional electric field parallel to the magnetic one, as well as
adding the Hartmann potential, etc. (A harmonic part is always necessary, though, due to the magnetic field).


Our strategy has been to start
 with the \emph{pure Kepler problem}  \cite{Cordani} 
 and then inquire what potential can be added such that separability in 
 (semi)parabolic coordinates is preserved. In the same spirit, we viewed the ``Runge-Lenz-type" conserved quantity $K_z$ in (\ref{QDRLpar})
as the Keplerian expression [represented by the
first and the third terms], ``corrected'' by the third one due to the oscillator.  

But we could have also started  at the other end, i.e.,  with the \emph{pure anisotropic oscillator}, which is separable, for $2:1$ ratio of the frequencies, in both Cartesian and (semi)parabolic coordinates  \cite{Makar,Boyer}. Then  we could have observed that  separability in (semi)parabolic coordinates is consistent with a Kepler potential of arbitrary strength, viewed as a perturbation of our initial oscillator. We could also view (\ref{QDRLpar})  as the conserved quantity
 related to \emph{oscillator-separability} [represented by the first and the third terms], ``corrected'' by the middle one, required due to the Keplerian perturbation. 
We mention that our problem here
can further be generalized by including magnetic charges \cite{Krivonos}.

\kikezd{Note added}
After this paper has been accepted, we received a message from J-W van Holten \cite{vanHolten}, pointing out that our results can also be derived using the  covariant framework of Ref. \cite{Killing} based on Killing tensors. Our conserved quantity (\ref{half3DRL}) is indeed associated to a \textit{fourth-rank Killing tensor} -- the only previously known examples being those discussed in Ref. \cite{4Killing}.

\begin{acknowledgments}
We are indebted to B. Cordani for his advice at the early stages of this project.
PAH acknowledges hospitality at the
 \textit{
Institute of Modern Physics in Lanzhou of
the Chinese Academy of Sciences}, and also
thank  V. Komornik and J-W van Holten for correspondence. 
 This work has been partially supported by the National Natural Science Foundation of
China (Grants No. and 11175215) and by the Chinese Academy of Sciences visiting
professorship for senior international scientists (Grant No. 2010TIJ06).
\end{acknowledgments}
\goodbreak


\end{document}